\documentstyle[11pt,epsf,paspconf,psfig]{article}

\begin{document}

\title{Four New BL\,Lac Surveys: Sampling New Populations}

\author{S. A. Laurent-Muehleisen, R. H. Becker}
\affil{IGPP/LLNL \& University of California-Davis, 7000 East Ave., Livermore,
CA, 94550, USA}

\author{W. Brinkmann, J. Siebert}
\affil{MPE, Postfach 1603, 85740 Garching, Germany}

\author{E. D. Feigelson}
\affil{The Pennsylvania State University, University Park, PA, 16802, USA}

\author{R. I. Kollgaard}
\affil{Fermi National Accelerator Lab, Batavia, IL, 60510, USA}

\author{G. D. Schmidt}
\affil{University of Arizona, Steward Observatory, Tucson, AZ, 85721, USA}

\author{P. S. Smith}
\affil{Kitt Peak National Observatory, Tucson, AZ, 85726, USA}

\begin{abstract}
The advent of large area deep radio and X-ray surveys is leading to the
creation of many new BL\,Lac samples.  In particular, the ROSAT All-Sky, Green
Bank and FIRST surveys are proving to be rich sources of new BL\,Lacs.  We
will discuss the methods used in four independent BL\,Lac searches based on
these surveys.  Comparison of the broadband spectral energy distributions of
these BL\,Lacs with those of previously known objects clearly points to the
existence of a large previously unrecognized population of objects with
characteristics intermediate between those exhibited by Low and High energy
peaked BL\,Lacs.
\end{abstract}

\keywords{X-ray emission, radio emission, BL Lac objects, surveys}

\section{Introduction}
The two best-studied samples of BL\,Lacs are the Einstein Extended Medium
Sensitivity X-ray and the 1\,Jy radio samples (Stocke et al.\ 1991; Stickel et
al.\ 1991).  However, numerous others exist including the Einstein Slew, HEAO,
EXOSAT, S4 and PG samples (Perlman et al.\ 1996; Wood et al.\ 1984; Giommi et
al.\ 1991, K\"{u}hr \& Schmidt 1990, Fleming et al.\ 1993) in addition to new
samples based on the ROSAT All-Sky Survey (see, e.g., Beckmann et al., these
proceedings).  With so many existing samples, why are more required?  First,
despite these attempts only $\sim$250 BL\,Lacs were known prior to the ROSAT
mission, and the largest complete samples consisted of only $\sim$30 objects.
Second, these objects were chosen with a variety of selection methods some of
which have biased previous samples against particular BL\,Lac subclasses.  New
samples which incorporate the observational consequnces of the underlying
physics which drives the BL\,Lac phenomenon are required if we are to
understand in detail the relationships between BL\,Lacs and other AGN or
between the different BL\,Lac subclasses.  Third, previous surveys have only
sampled the brightest objects in either the X-ray or radio bands which has
led to gaps of two orders of magnitude between, e.g., the radio fluxes of
sources in radio- and X-ray-selected samples.  Interesting populations of
objects lurk between the extremes of radio- and X-ray-dominated objects and
discovery of them requires delving into fainter populations.  Finally, only
now are suitable parent surveys available which will allow us to satisfy these
goals.

X-ray surveys contain relatively large fractions of BL\,Lacs.  Coupled with
the fact that radio-silent BL\,Lacs either do not exist or are exceedingly
rare (Stocke et al.\ 1990), correlation of X-ray and radio surveys is an
efficient means of creating new BL\,Lac samples.  Therefore, we have chosen
the ROSAT All-Sky Survey (RASS), the 6\,cm Green Bank radio (Gregory et al.\
1996) and FIRST 20\,cm surveys (Becker et al.\ 1995) for our parent surveys.

\section{RASS-Green Bank}
The RGB sample consists of 1567 sources found in a correlation of the RASS and
Green Bank catalogs.  Optical spectroscopy of previously unidentified objects
revealed 38 new BL\,Lacs (Laurent-Muehleisen et al.\ 1998) and resulted in a
final sample of 125 RGB BL\,Lacs.  These objects span a large range in radio
(3$<$S$_{\rm r}$$<$2160\,mJy), optical (13.3$<$B$<$21.2) and X-ray (3$\times
10^{-13}$$<$F$_{\rm x}$$<$$4\times 10^{-10}$erg s$^{-1}$cm$^{-2}$) fluxes.
Optical polarimetry shows that $\sim$60\% of the sources are polarized $>$3\%
(Pursimo et al., these proceedings).  The RGB Complete sample consists of 32
BL\,Lacs with B$\le$18.0\,mag in a 3970\,deg$^2$ region of sky.  Figures 1 and
2 show the broadband spectral energy distribution of the RGB BL\,Lacs.  These
figures clearly indicate that intermediate BL\,Lacs, those between the
previously disparate classes of Low and High energy peaked BL\,Lacs (LBLs and
HBLs, respectively) are present in large numbers and that no clear division
exists between the two subclasses (see also Siebert et al., these
proceedings).

\begin{figure}
\plottwo{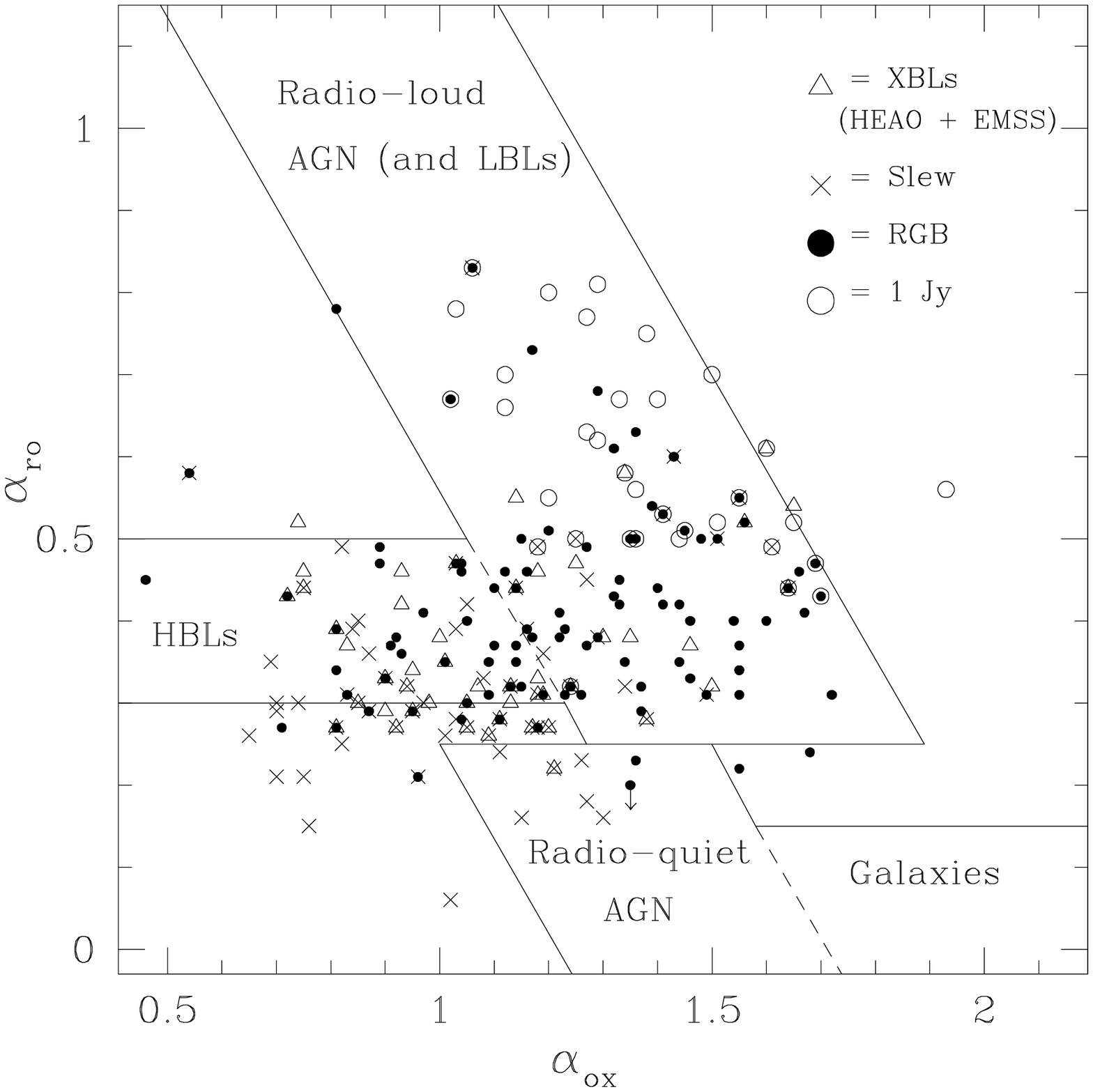}{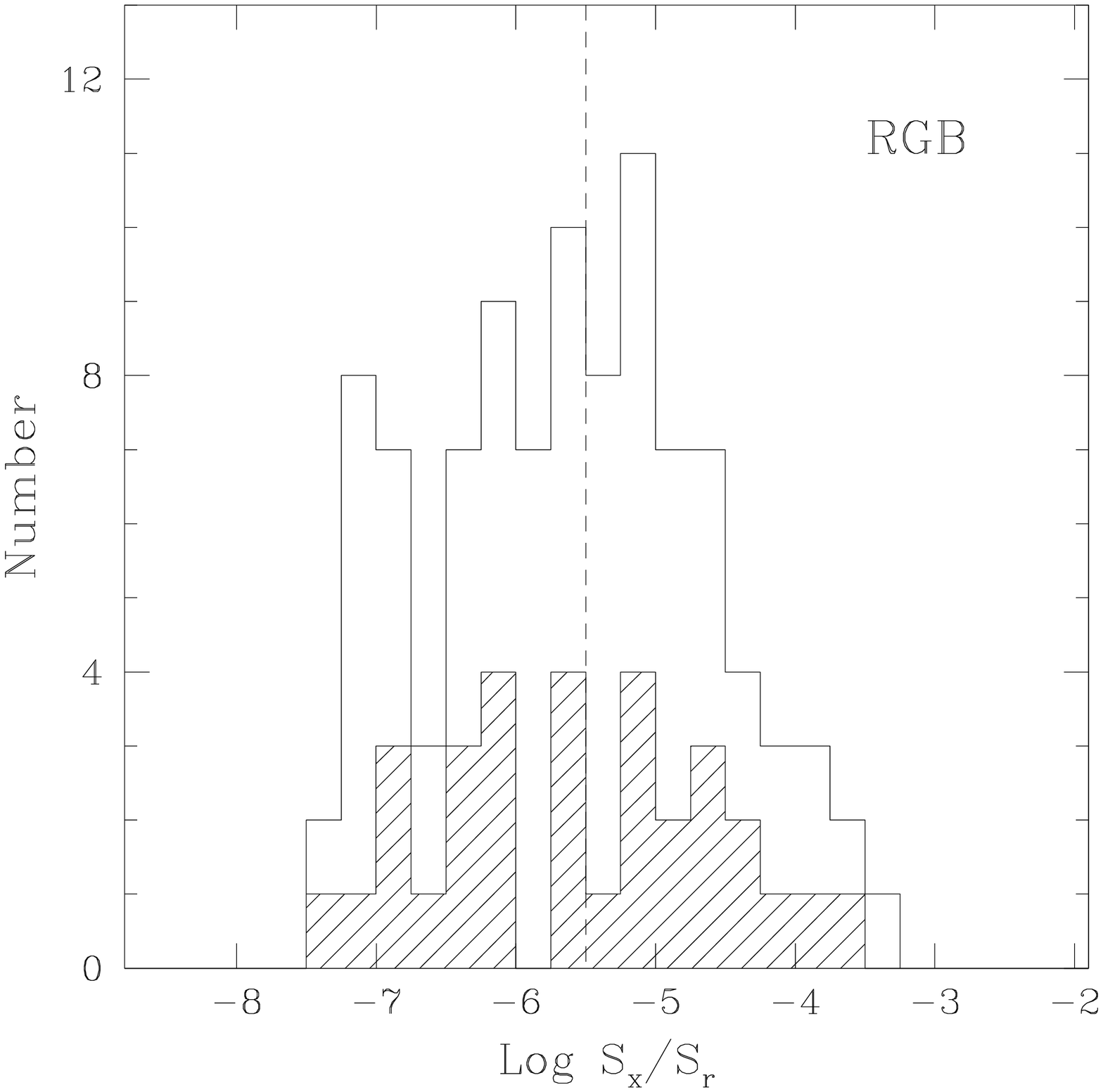}
\caption{(left) The $\alpha_{\rm ro}$ vs.\ $\alpha_{\rm ox}$ spectral index
distribution for the RGB and comparison samples.  The RGB sample clearly spans
the traditional HBL and LBL regimes in addition to the intermediate region.}
\vspace*{-0.3cm}
\caption{(right) The distribution of $\log$(S$_{\rm x}$/S$_{\rm r}$) for the
RGB BL\,Lacs.  The shaded histogram denotes the distribution of RGB Complete
sample. The vertical line indicates the canonical HBL/LBL division.}
\end{figure}

\section{RASS-FIRST}
The biggest advantage the RASS-FIRST correlation has compared to the RGB
survey is its twenty times deeper radio flux limit which enables the study of
lower radio luminosity sources and also moderate luminosity sources to higher
redshifts.  This facilitates a better study of the effects of cosmic evolution
vs.\ beaming on BL\,Lac LogN-LogS distributions and luminosity functions.  The
RASS-FIRST sample currently consists of 82 BL\,Lacs in a 3000\,deg$^2$ region
of the sky.  Spectroscopic identifications are continuing but the BL\,Lacs
found to date range from 1\,mJy to 1500\,mJy and from very low redshift
(0.034) to z$=$1.37.

\section{FBQS BL\,Lac Sample}
The FIRST Bright Quasar Survey (FBQS; Gregg et al.\ 1996; White et al.\ 1998)
is a program designed to identify a new deep radio-selected sample of quasars.
Spectroscopy of bright (R$<$17.8), blue (B$-$R$<$2.0) optically
unresolved sources has shown that this survey is also a surprisingly rich
source of BL\,Lacs, yielding a sample of 158 objects.  The optical color and
morphology criteria have undoubtedly biased the sample, but it has the
advantage, unlike the RGB or RASS-FIRST samples, of selecting BL\,Lacs
independent of their X-ray emission.  It is therefore most similar to the
radio-selected 1\,Jy sample, but with a flux limit 1000 times more sensitive.
This sample will undoubtedly be useful for studying the transition between
LBLs and HBLs at radio flux densities between 1 and 1000\,mJy, a regime which
presently is dominated by X-ray-selected samples.

\section{FIRST Flat Spectrum Sample}
BL\,Lacs, like all blazars, exhibit flat radio spectral indices, a fact used
to create the 1\,Jy BL\,Lac sample.  The high radio flux limit of the 1\,Jy
sample can easily be improved upon using the FIRST and Green Bank catalogs.
We have constructed a sample of $\sim$5000 sources with radio spectral indices
flatter than 0.5 between 6 and 20\,cm and brighter than 35\,mJy.  We have
obtained spectroscopic identifications for $\sim$1/3 of the 1600 sources with
counterparts on the POSS I plates and constructed a sample of 68 BL\,Lacs.
Like the FBQS BL\,Lac sample, the FIRST Flat Spectrum sample is selected
completely independent of X-ray emission.  This sample will be ideal for
studying the transition between Flat Spectrum Radio Quasars and BL\,Lacs.

\section{Summary}
New BL\,Lac samples based on the Green Bank and FIRST radio surveys and the
RASS X-ray survey are proving highly successful.  In Figure 3 we show the
radio LogN-LogS distribution for all known BL\,Lacs in the FIRST survey (See
\S 3-5).  Clearly these samples bridge the gap between HBLs and LBLs and will
provide useful databases for constraining unification models and the
relationship between BL\,Lacs and other classes of AGN.

\begin{figure}
\psfig{figure=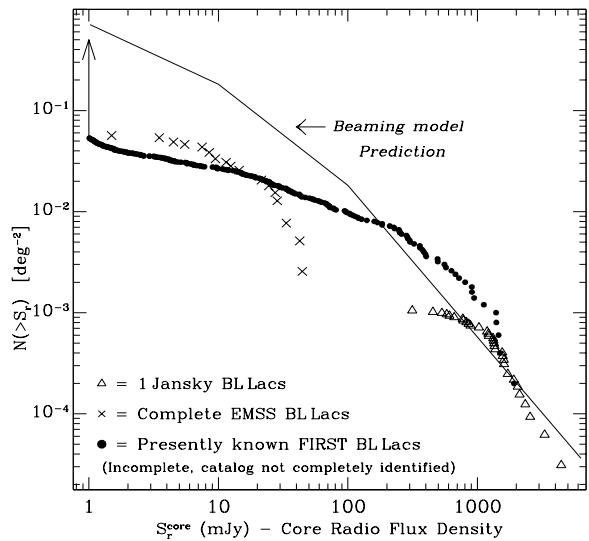,height=3.0in,width=3.2in}
\caption{The LogN-LogS distribution for all presently known BL\,Lacs in the
FIRST survey.}
\end{figure}

\end{document}